\documentclass[prd,twocolumn,superscriptaddress,floatfix,amsmath,amssymb]{revtex4-1}
\usepackage[english]{babel}
\usepackage{graphicx}
\usepackage{dcolumn}
\usepackage{bm}
\usepackage{verbatim}
\usepackage{mathrsfs}
\usepackage{cancel}
\usepackage{color}
\newcommand{\ket}[1]{\left| {#1} \right\rangle}

\newcommand{\proj}[2]{\left| {#1} \right\rangle\!\left\langle {#2} \right|}

\newcommand{\ii}{\mathrm{i}}

\newcommand{\tr}{\operatorname{Tr}}

\def\slashchar#1{\setbox0=\hbox{$#1$} 
\dimen0=\wd0 
\setbox1=\hbox{/} \dimen1=\wd1 
\ifdim\dimen0>\dimen1 
\rlap{\hbox to \dimen0{\hfil/\hfil}} 
#1 
\else 
\rlap{\hbox to \dimen1{\hfil$#1$\hfil}} 
/ 
\fi}

\begin{document}

\title{Processing quantum information with relativistic motion of atoms}
%

\author{Eduardo Mart\'{i}n-Mart\'{i}nez}
\affiliation{Institute for Quantum Computing and Department of Physics and Astronomy, University of Waterloo, 200 University
Avenue W, Waterloo, Ontario, N2L 3G1, Canada}
\affiliation{Dept. Applied Math., University of Waterloo, 200 University
Av W, Waterloo, Ontario, N2L 3G1, Canada}
\affiliation{Perimeter Institute for Theoretical Physics, Waterloo, Ontario N2L 2Y5, Canada}
\author{David Aasen}
\affiliation{Dept. Applied Math., University of Waterloo, 200 University
Av W, Waterloo, Ontario, N2L 3G1, Canada}
\author{Achim Kempf}
\affiliation{Dept. Applied Math., University of Waterloo, 200 University
Av W, Waterloo, Ontario, N2L 3G1, Canada}
\affiliation{Centre for Quantum Computing Technology, Department of Physics, University of Queensland, St. Lucia, Queensland 4072, Australia}

\begin{abstract}
We show that particle detectors, such as 2-level atoms, in non-inertial motion (or in gravitational fields) could be used to build quantum gates for the processing of quantum information. Concretely, we show that through suitably chosen non-inertial trajectories of the detectors the interaction Hamiltonian's time dependence can be modulated to yield arbitrary rotations in the Bloch sphere due to relativistic quantum effects.\end{abstract}

\maketitle

{\it Introduction.-} The study of the interface between general relativity and quantum theory has long been a fruitful area of research which, more recently, has included the use of quantum information techniques. In this letter, we explore the idea that gravity or non-inertial motion can be used, in principle, to aid quantum information processing and computing.

The concept of using non-inertial motion to perform quantum computing has been suggested by I. Fuentes. Concretely, Fuentes and collaborators showed that by accelerating optical cavities in relativistic regimes they can perform two-mode squeezing transformations on the field modes inside the cavity  \cite{IvetteRef,Prdvette}. Our aim  is different in that we consider accelerating detectors instead of field mode transformations, i.e., we consider for example the acceleration of atoms (as an instance of detectors that encode qubits in their internal states) instead of moving cavities. 

It is known that acceleration induces squeezing in the field modes \cite{Unruh1976} and that this can lead to entanglement amplification effects \cite{MigC,CIvyty3}. However, studies of models of accelerated atomic detectors have always found that acceleration (and gravity) act as a source of noise, thereby degrading entanglement and quantum correlations \cite{matsako,matsako2}.

In view of the previous literature one might therefore expect that to accelerate detectors that carry qubits, such as atoms, could not be beneficial for the processing of the quantum information in those qubits. Here, we will show that in contrast to this naive expectation,  intense gravity or the non-inertial motion of accelerated atoms in a cavity can indeed be used to build arbitrary quantum gates that act on the internal state of the atom: we can control the interaction as a function of time by controlling the relativistic motion of the atoms and by using the associated effects related to relativistic time dilation and length contraction. Thus, the key finding is that control over the acceleration of atoms can be used to perform quantum information tasks as a direct consequence of general relativistic quantum effects. 

To this end, we will first show how acceleration can induce controlled motion in the Bloch sphere of a qubit's internal state. As an illustrative example we will consider the interaction of the atom with the vacuum state of the field in a cavity. Then, we will show that by making atoms accelerate while they interact with a coherent state in a cavity, arbitrary 1-qubit quantum gates can be implemented. We will show how and why similar results cannot be obtained with inertial settings. We will finally discuss the consequences these findings may have regarding the detection of general relativistic quantum effects and we will discuss the prospects for experimental implementations. We will see that the accelerations required are achievable in principle with state-of-the-art  particle acceleration technology. Furtherfore, we will discuss that a quantum analogue of our scenario can realistically be implemented as a simulation in trapped ion systems or superconducting circuit setups.

{\it The setting.-} We consider a 2-level atom as our qubit. This system will be coupled to the quantum field inside a cavity. The interaction of an atom and the radiation inside a cavity can be very well approximated (for atomic transitions with no exchange of angular momentum) by the Unruh-Dewitt Hamiltonian, as shown in  \cite{Wavepackets}. This Hamiltonian models the interaction of a two-level system with a scalar field \cite{DeWitt}. The Hamiltonian is $H=H_0^{(\text{d})}+H_0^{(\text{f})}+H_I$, being $H_0^{(\text{d})}$ and $H_0^{(\text{f})}$ respectively the  free Hamiltonian of  the two-level system and the field and $H_{I}$ the interaction Hamiltonian $
H_{I}= \lambda \ \xi(\tau) \mu(\tau) \phi(x(\tau)),$
where $\lambda$ is the coupling strength, $\xi(\tau)$ is a switching function controlling the interaction time, $\mu(\tau)$ the monopole moment operator and $x(\tau)$ the worldline of the atom. $\Omega$ will be the energy difference between the ground and excited state of the 2-level system (we will refer to it as the `detector', the 'atom' or the `qubit'). The detector is coupled through its monopole moment to the massless scalar field $\phi (x)$  along its worldline. In the interaction picture $H_I$ takes the form
\begin{equation}\label{hamilto}
H_{I} =  \lambda\, \mu(\tau) \sum_{j=1}^{\infty} (a_{j}^{\dagger}e^{i \omega_{j} t(\tau)}+a_{j} e^{-i \omega_{j} t(\tau)})\sin{k_j x(\tau)},
\end{equation}
where the monopole moment of the qubit takes the usual form $\mu(\tau)=\sigma^{+}e^{i \Omega \tau}+\sigma^{-}e^{-i \Omega \tau}$. This Hamiltonian is essentially equivalent to the infinitely multimode Jaynes-Cummings model.

Notice that we have chosen to expand the field in terms of the standard basis of Minkowskian stationery waves. Depending on the detector's trajectory, the relationship between the Minkowskian time and position $(t,x)$ and the proper time of the detector $\tau$ will vary. In the simplest scenario of a stationary inertial atom the worldline of the detector is given by $x=x_0$ and $t=\tau$. In the still relatively simple case of a uniformly accelerated detector of fixed acceleration $a$, $t$ and $x$ are parametrized in terms of the proper time of the detector $\tau$ as
\begin{equation}\label{Rindler}
x(\tau) =a^{-1} (\cosh{a\tau}-1),\qquad t(\tau) = a^{-1}\sinh{a\tau},
\end{equation}
where we have assumed for simplicity $c=1$ and that at $\tau=t=0$ the detector is in the $x=0$ position of the cavity. Hence, this leaves us with a time-dependent Hamiltonian $H(\tau)$.

The time evolution under this Hamiltonian from a time $\tau=0$ to a time $\tau=T$ is given by the following expansion
\begin{equation}\label{pert}
U(T,0) = \openone\!-\ii\int_{0}^{T}\!\!\!\!\!d \tau H_{I}(\tau) - \int_{0}^{T}\!\!\!\!\!d\tau \!\!\int_{0}^{\tau}\!\!\!\!\!d\tau_{1}\,    H_{I}(\tau) H_{I}(\tau_{1})+\hdots
\end{equation}
Under the realistic assumption that the coupling strength is small enough, we can neglect higher orders in \eqref{pert}. If we denote by $\rho_0$ the initial density matrix of the field-detector system we get that after a time $T$, $\rho_T=\rho_0+\rho_T^{(1)}+\rho_T^{(2)}+\mathcal{O}(\lambda^3)$, where
\begin{align}
\label{e1}\rho_T^{(1)}&=U^{(1)}\rho_0+\rho_0{U^{(1)}}^\dagger\\
\label{e2}\rho_T^{(2)}&=U^{(1)}\rho_0{U^{(1)}}^\dagger+U^{(2)}\rho_0+\rho_0{U^{(2)}}^\dagger.
\end{align}

{\it Accelerated detector in Vacuum.-} Prior to showing how to implement arbitrary Bloch sphere rotations based on relativistic motion we will first analyze one of the advantages of controlling the interaction:
It is well known that, if we prepare the vacuum state in a cavity, an inertially moving atom will not be able to reach every point of the Bloch sphere no matter how much time of evolution we allow. As an example of this, an atom in the ground state will never evolve into the excited state if there are no photons to absorb. More specifically, the reason is that the terms in the Hamiltonian \eqref{hamilto} that would allow such transitions (the counter-rotating terms whose nature we will review below) are suppressed  for non-negligible times due to their highly oscillatory nature (See for instance \cite{ScullyBook}).

However, we will discuss below that this is not the case for an accelerated detector. Due to relativistic effects the rotating and counter-rotating terms both become equally important for the relevant timescales.

For instance, an accelerated Unruh-DeWitt detector probing the vacuum state of the field would detect instead a distribution of field quanta \cite{Birrell1982} due to the contribution of the counter-rotating terms (the celebrated Unruh effect). It is therefore not surprising that,  even in the vacuum, an atom can non-trivially move around the Bloch sphere by controlling its acceleration. Although using the vacuum state is not optimal  to show that arbitrary 1-qubit gates can be implemented, it  constitutes a first example to illustrate the differences between the inertial and the accelerated case, and we will briefly analyse it prior to showing how to perform arbitrary rotations in the Bloch sphere: Directly from \eqref{pert}, the first order contribution to the evolution operator is given by
\begin{align}
&\nonumber U^{(1)}\!\!=\!\!\frac{\lambda}{\ii}\sum_j \big(\sigma^+ a_j^\dagger I_{+,j}\!+\!\sigma^- a_j I_{+,j}^*\!+\!\sigma^- a^\dagger_j I_{-,j}\!+\!\sigma^+ a_j I_{-,j}^*\big)\\
\label{integro}
&I_{\pm,j} \equiv I_{\pm,j}(T)=   \int_{0}^{T} d\tau e^{\ii[\pm \Omega \tau + \omega_j t(\tau)]} \sin{[k_j x(\tau)]}.
\end{align}
For an inertial detector this is the well known integration of the rotating and counter-rotating terms with the typical resonance condition $\omega_j=\Omega$. However in the accelerated case, after substituting $x(\tau)$ 
 and $t(\tau)$ with the parametrization \eqref{Rindler} we observe that the phases depend very nontrivially on time, so that 1) the resonance condition is time dependent  and 2) the counter-rotating terms become comparable to the rotating ones very quickly.

Let us now begin with an arbitrary state for the qubit and the vacuum  in all the modes of the cavity, namely $\rho_0=\rho_{0,{(\text{d})}}\otimes_{j}\proj{0_j}{0_j}$. By simple inspection of $U^{(1)}$ one can readily check that $\rho_{T,(\text{d})}^{(1)}=\tr_{(\text{f})}\rho_T^{(1)}=0$. Therefore the leading contribution to the qubit evolution comes from the second order (in $\lambda$) density matrix perturbation. Since the field is originally in the vacuum state, after some lengthy computations one finds that
 \begin{figure*}[t] \begin{tabular}{cc}
\!\includegraphics[width=0.61\textwidth]{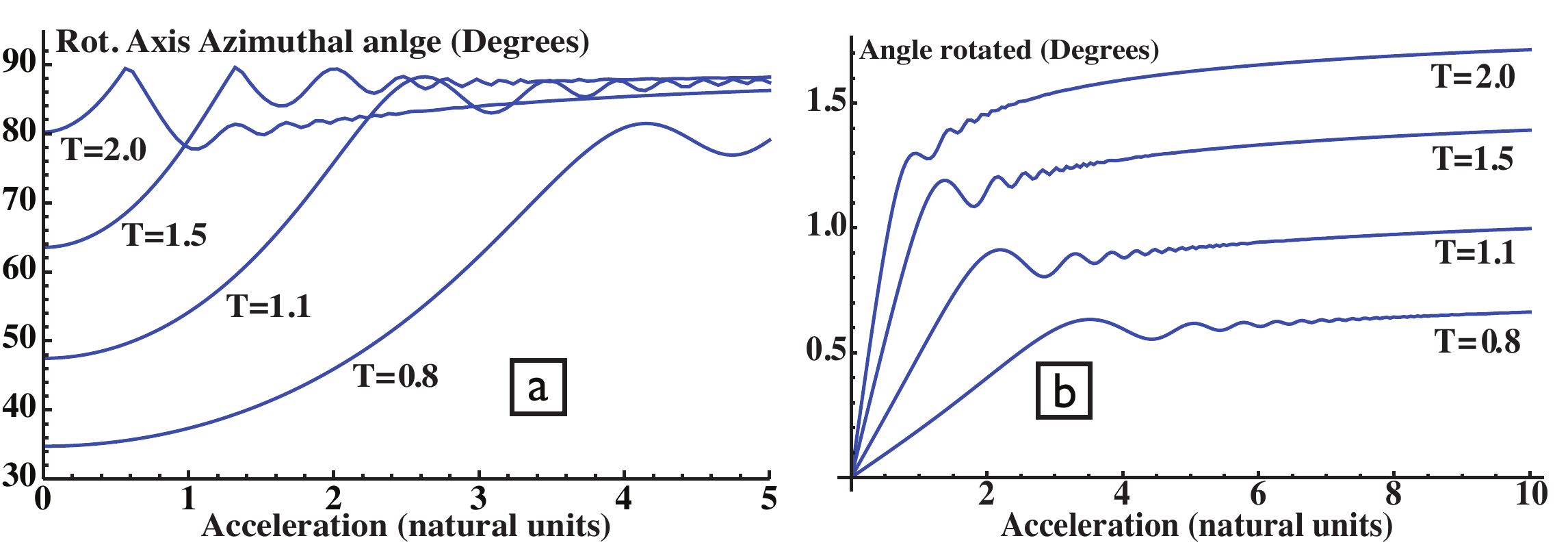}&
\includegraphics[width=0.39\textwidth]{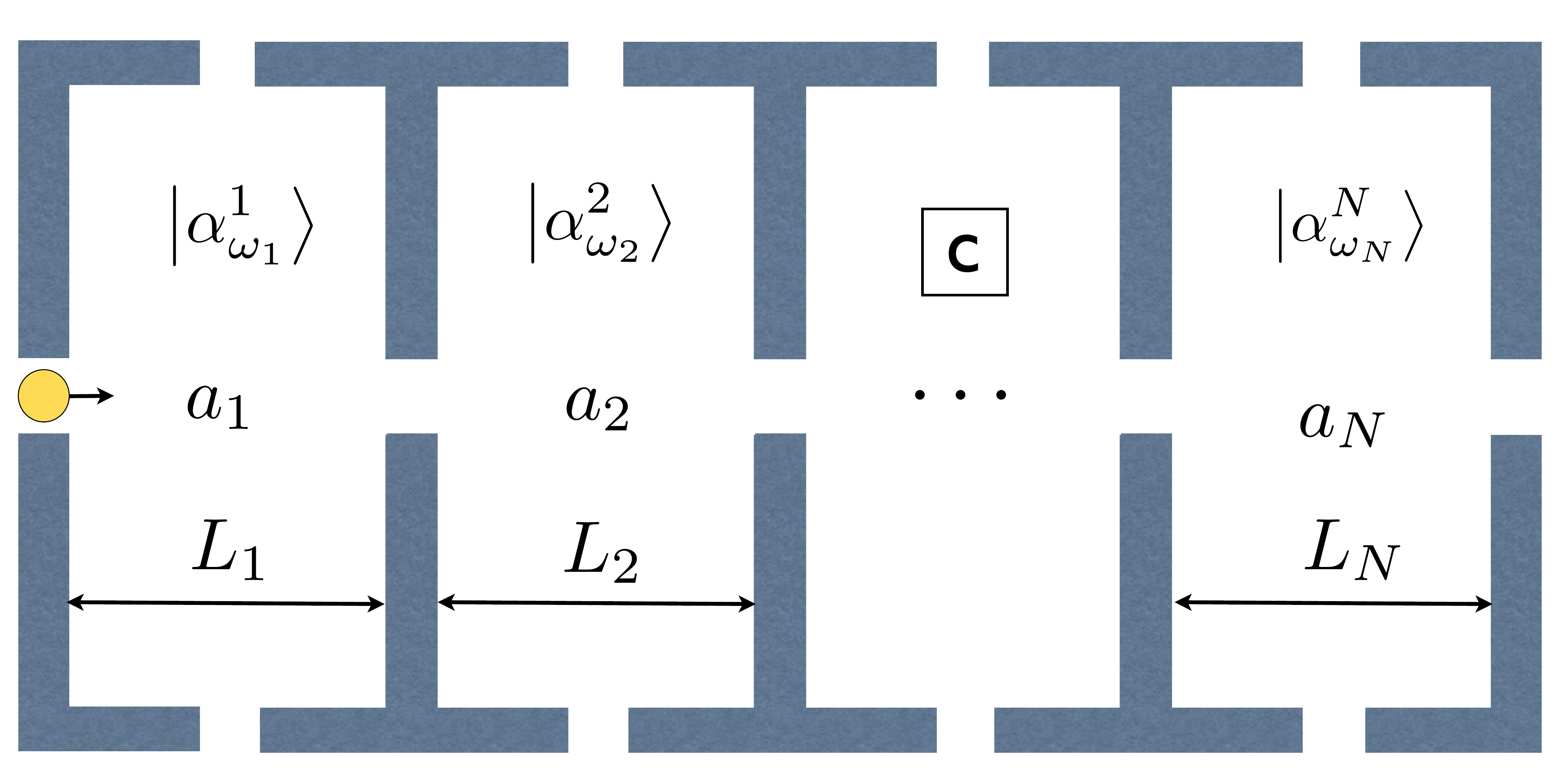}\end{tabular}
  \caption{For $\lambda|\alpha|=0.01$, {\bf a)} Azimuthal angle of the rotation axis for various $T$.  {\bf b)} Magnitude of the rotation $\delta$ for these cases. Independent rotations that are more than 50 degrees apart can be achieved by controlling the atom's acceleration. $\Omega$ fixes the timescale and the units of $T$. Acceleration is expressed in natural units (we took $c=1$), so it has units of inverse time. {\bf c)} Scheme of an array of cavities with pre-prepared coherent states $ \ket{\alpha^{1}}_{\omega_1}\bigotimes_{\omega\neq\omega_{1}}\ket{0_\omega}$ that allow successive different rotations in the Bloch sphere, accelerations could alternate sign.}
  \label{fig:results}
\end{figure*}
\begin{align*}
  \nonumber\tr_{(\text{f})}\!\!\big[U^{(1)}&\!\rho_0{U^{(1)}}^\dagger\big]\!=\!\lambda^2\!\sum_{j=1}^\infty\!\big(A^{x}_j\sigma_{x} \rho_{0,(\text{d})} \sigma_{x}\!+\!A^{y}_j\sigma_{y} \rho_{0,(\text{d})} \sigma_{y}\\
 & +\!A^{xy}_j\sigma_{x} \rho_{0,(\text{d})} \sigma_{y}+A^{yx}_j\sigma_{y} \rho_{0,(\text{d})} \sigma_{x}\big),\\[3mm]
\nonumber A^{x/y}_j&= |I_{+,j}|^{2}+|I_{-,j}|^{2}\pm I_{+,j}^{*}I_{-,j}\pm I_{+,j}I_{-,j}^{*},\\
\nonumber  A^{xy}_j&=\ii  [(I_{+,j}I_{-,j}^{*}-I_{+,j}^{*}I_{-,j})-|I_{+,j}|^{2}+|I_{-,j}|^{2}]
\end{align*}
and $A^{yx}_j=\big(A^{xy}_j\big)^*$. The derivation of the contribution coming from the second term in \eqref{e2} is more involved but can be simplified given that we are interested only in $\tr_{F}(U^{(2)}\rho_0)$ with the field initially in the vacuum state:
\begin{align}
\nonumber&\tr_{F}(U^{(2)}\rho_0) = \lambda^2 \sum_{j=1}^{\infty}\left(  C_{\openone,j} \rho_{D0} +C_{z,j} \sigma_z \rho_{D0}  \right)\\
&\nonumber C_{\openone,j} = 2(M_{j,-,+}+M_{j,+,-}),\;C_{z,j} = 2(M_{j,-,+}-M_{j,+,-})\\
&M_{j,\pm,\pm^{\prime}} = \int_{0}^{T} d \tau\, I_{\pm^{\prime},j}(\tau)  \frac{d}{d\tau} I_{\pm,j}(\tau)^{*} .
\end{align}
We can now compute the change of the Bloch vector $\bm b=(b_x,b_y,b_z)$ after time evolution. If we define $B_{\pm,j}=I_{+,j}I_{-,j}^{*}\pm I_{+,j}^{*}I_{-,j},\; D_{\pm,j}=|I_{+j}|^2\pm|I_{-j}|^2$ then
  \begin{align}
\nonumber \Delta b_{x} &\!=\! 2\lambda^2 \!\sum_{j}\!\big[ \big(\frac{B_{+,j}}{4}\!+\!\operatorname{Re}C_{\openone,j}\big) b_{x} \!+\!  \big(\frac{\ii B_{-,j}}{4}\!+\!\operatorname{Im}C_{z,j}\big) b_{y} \big]\\
\nonumber \Delta b_{y} &\!= \!2\lambda^2 \sum_{j}\! \big[\big(\operatorname{Re}C_{\openone,j}\!-\!\frac{B_{+,j}}{4}\big)b_{y}\!+\!  \big(\frac{\ii B_{-,j}}{4}\!-\!\operatorname{Im}C_{z,j}\big)b_{x}\big] \\
 \Delta b_{z} &= 2\lambda^2 \sum_{j} \big[\big(\operatorname{Re}C_{\openone,j}\!-\!\nonumber \frac{D_{+,j}}{4}\big)b_z +\frac{D_{-,j}}{4} \!+\! \operatorname{Re}C_{z,j} \big].
\end{align}
One can check  that the internal state of an inertial detector cannot be moved towards the north pole of the Bloch sphere when it is in the northern hemisphere. However in the accelerated case it is possible to move in any direction of the Bloch sphere by just increasing the acceleration, which makes the rotating and counter-rotating terms comparable. Also, the acceleration introduces a dynamical Doppler effect that gets the atom in resonance with several modes during its time inside the cavity, making the atom resonate successively to several different modes of the field (multiple level-crossing \cite{Diegger}).

This setting is, however, not very useful to achieve universal 1-qubit gates since the state of the atom gets mixed at the same order of perturbation theory as the rotation effects appear. We will now discuss how to perform arbitrary 1-qubit quantum operations on the internal state of the atom by preparing a coherent state in the cavity, which is also simpler to prepare than the vacuum.

{\it Universal 1-qubit gates with coherent states.-} In this section, we show the main claim of the letter: that one can achieve arbitrary rotations by preparing coherent states in one of the modes of the cavity. Let us consider that now the initial state of the system is $
\rho_0=\rho_{0,(\text{d})}\otimes\proj{\alpha_{\omega_i}}{\alpha_{\omega_i}}\bigotimes_{j\neq i}\proj{0_{\omega_j}}{0_{\omega_j}}.$

In this case the leading order is given by \eqref{e1}. In fact, if the rest of the modes are not populated, their contributions will be of second order in $\lambda$. To compute the leading order time evolution we need to calculate $\tr_{\text{f}}(U^{(1)}\rho_0)$. This is particularly simple given that
$\tr_{\text{f}}(a\rho_0)=\alpha \rho_{0,{(\text{d})}}$ and $\tr_{\text{f}}(a^{\dagger}\rho_0)=\alpha^* \rho_{0,{(\text{d})}}$. Hence, we have
 \begin{equation}
\nonumber\tr_{\text{f}}(U^{(1)}\!\rho_{0}) \!= \!\frac{\lambda}{\ii}\!\! \left[\alpha(\sigma_{+} I_{-}^{*}\! +\! \sigma_{-} I_{+}^{*})\!+\!\alpha^*\!( \sigma_{+} I_{+} \!+\!\sigma_{-} I_{-})  \right] \!\rho_{0,(\text{d})}
\end{equation}
where $I_{\pm}$ is the integral \eqref{integro} for the mode $i$ where the coherent state is prepared. Defining $A = \alpha^* I_{+} + \alpha I_{-}^*$,
\begin{equation}\label{coherent}
\rho_{T,(\text{d})} \!=\! \rho_{0,(\text{d})} + \frac{\lambda}{\ii} \!\left[(A\!+\!\!A^*)\sigma_{x} \rho_{0,(\text{d})} \!+\! \ii(A\!-\!\!A^*)\sigma_y \rho_{0,(\text{d})}\! \!-\!\text{H.c.} \right]
\end{equation}

An infinitesimal rotation on the Bloch sphere of angle $\delta$ around the axis defined by $\bm {n}$ is $
R_{\bm{n}}(\delta) \approx  \openone-\ii \frac{\delta}{2}(\bm{n}\cdot \bm{\sigma})
$. Its action on a density matrix $\rho_{0,(\text{d})}$ would be
$
R_{n}(\delta)\rho_{0,(\text{d})}R_{n}(\delta)^{\dagger} \approx \rho_{0,(\text{d})} - \ii \frac{\delta}{2} (\bm{n} \cdot \bm{\sigma} \rho_{0,(\text{d})} - \text{H.c.}).
$

By inspection we see that  \eqref{coherent} has the form of an infinitesimal rotation  around the axis defined by the direction of the (unnormalized) vector
\begin{equation}
\bm{n} = (A+A^*, \ii(A-A^*),0),
\end{equation}
and the magnitude of the rotation  is 
\begin{equation}
\delta =2 \lambda |\bm{n}|=4\lambda|\alpha| | e^{-\ii\operatorname{Arg}\alpha} I_{+} +e^{\ii\operatorname{Arg}\alpha} I_{-}^*|.
\end{equation}
 
Therefore we can perform unitary rotations thus introducing no mixedness at leading order in the coupling strength. For the perturbative calculation to be valid, we require that $\lambda|\alpha|\ll1$ (See \footnote{We also require $T$ not to be arbitrarily large. For the regimes studied here it is safe to consider $T\lesssim100$}). If this is fulfilled any non-unitarity of the transformation coming from truncating the perturbative series (and therefore the introduced mixedness) would be negligible. Note that the rotation axis is independent of $\lambda$ and therefore we would be able to vary it regardless of how small the coupling strength.

For an accelerated atom, $A=A(a,T)$ and $\bm n=\bm n(a,T)$ are functions of acceleration and interaction time. Controlling the atom's acceleration and the interaction time we can control both the axis with respect the rotation is performed and the magnitude of the rotation.

\begin{figure} 
\includegraphics[width=0.45\textwidth]{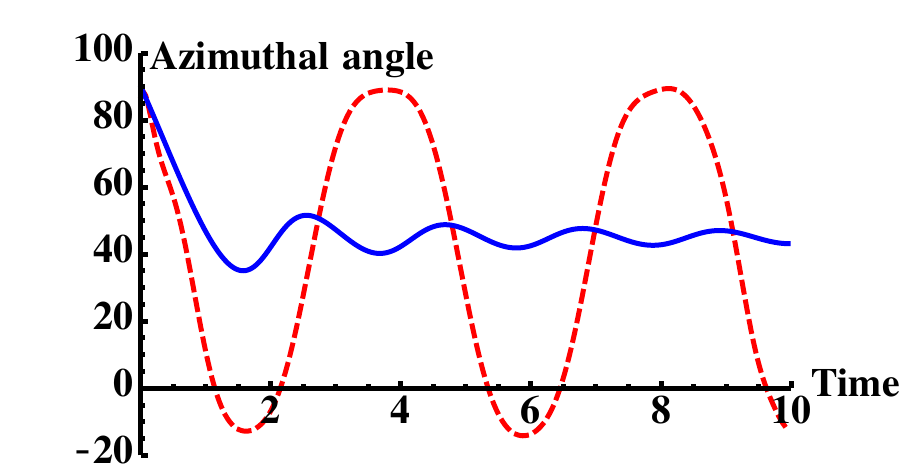}
  \caption{For $\lambda|\alpha|=0.01$,  variation in the rotation axis  as a function of time for an atom at rest centred in the cavity (Blue solid) and and an accelerated atom with fixed acceleration $a=1$ (red dashed). For the inertial case the rotation axis quickly stagnates, whereas in the accelerated case we observe a much greater variation that does not get damped with time.}
  \label{fig:comparison}
\end{figure}

To be able to freely move in the Bloch sphere by means of several of these small rotations we need to prove that one can perform at least two independent rotations at every single point in the Bloch sphere. We see in Fig. \ref{fig:results}a the azimuthal angle of the rotation axis $\phi_{_T}(a)$ as a function of acceleration for a fixed interaction time, showing that  independent rotations can be achieved by varying acceleration. Fig. \ref{fig:results}b shows the magnitude of the rotation. In a similar fashion we can compute $\phi_a(T)$ to evaluate the variation  in the rotation axis as a function of T for a fixed acceleration. As portrayed in Fig. \ref{fig:comparison}, the rotation axes can be controlled to be even more than a hundred degrees apart  controlling the interaction time $T$.

We see that given the rotation axis dependence on $a$ we can make completely independent rotations in the Bloch sphere by controlling  the atom's acceleration. Out of the composition of such rotations, an arbitrary trajectory in the Bloch sphere can be tailored by letting the atom describe accelerated trajectories through an array of several cavities as shown in Fig. \ref{fig:results}c.

{\it Comparison with the inertial case.-}   Let us compare  the results above with an equivalent setting in which we have an inertial atom and assume that we have control over the total interaction time. Although  there is indeed some variation of the rotation axis when we increase $T$, we can show that such variation is always much smaller than what we would obtain for an equivalent setting but with a fixed acceleration. More importantly, in the inertial case as time increases, the rotation axis starts a damped precession around a fixed vector in the Bloch sphere, rendering the time controlling technique useless in order to perform rotations around different axes if the atom is not accelerated. This can be seen in Fig.  \ref{fig:comparison}.

{\it Experimental feasibility and two qubit gates.- } Let us consider the magnitudes involved in a possible experimental implementation. The natural scale of units is fixed by fixing units for $\Omega$, namely $\tilde a=a(\Omega c / \pi)$.  For atomic gaps of GHz, one natural unit of acceleration is equivalent to $10^{16}g$ (g is the Earth surface gravity) so to have non-trivial rotations we would need accelerations of $\sim10^{15}g$. This acceleration is one order of magnitude better than the best previous proposal for detection of the Unruh effect with the same atomic gap \cite{BerryPh} and plausible in theory \cite{ruso}. In fact, the acceleration required can be further reduced using a detector with a smaller gap. For example the use of hyperfine transitions or nuclear spin as our qubit  will reduce the energy gap to order of MHz  \cite {hyperfine1} thereby reducing the required accelerations to $\sim10^{12}g$. This is also the scale of accelerations that can be reached at the LHC \cite{Atlas1}. Additionally one can think of Stark shifted atomic levels or Zenner-induced transitions as qubits to achieve regimes of Hz hence reducing the accelerations to $\sim10^{6}g$. Those accelerations are indeed experimentally achievable for the short times required.

 Although theoretically within reach of current technology, an experimental realization would be much easier to achieve in analogue systems. Current technology of ion trapping and superconducting circuits already allows for experiments where relativistic effects can be observed \cite{supercond,DCasimir,PastFutPRL}. In particular, the simulation of relativistically accelerating atoms in trapped ion systems and superconducting circuits was formerly studied in \cite{Diegger}. The simulations proposed in \cite{Diegger} are precise analogues of the physical setting required here. 

Specifically, for the simulation in trapped ions, an analysis of orders of magnitude is also provided in \cite{Diegger}. A conservative estimation of the accelerations that could be reached in these simulations is $a/c\approx (10^{-3}-10^{-1})\Omega$, which already allows for the observation of all the effects studied here. As discussed in \cite{Diegger}, current technology in acusto-optical resonators can produce variations of the optical phase in a rate much beyond the required scales to build an experimental realization of what is proposed in this letter. This would be achieved  by means of standard experimental techniques from trapped ion quantum computation  \cite{Leibfried2003}.

Additionally, and as also discussed in \cite{Diegger}, superconducting qubits ultra-strongly coupled to a microwave cavity \cite{Wallraff} provide a natural setup of an analogue setting where this experiment can be realized. In this case, the relativistic atom Hamiltonian is simulated by means of  the driving of the qubit frequency using the techniques published in \cite{sidebands}.

Regarding two-qubit operations, there is always the possibility of performing only the single qubit operations through relativistic motion while performing the two-qubit operations through more traditional methods, such as, for instance, by trapping the previously accelerated ions and then applying conventional trapped ions techniques for implementing two-qubit gates. This is possible because arbitrary one qubit operations can be performed also under the constraint of the trajectory ending at rest.

 On the other hand, there is the exciting possibility that two-qubit gates could also be driven through  motion. 
Indeed,  as soon as there is an interaction between the two systems via the field, one generically obtains an entangling two-qubit unitary, as one can readily check \cite{Reznik2005,Olson2011,PastFutPRL}.  We can therefore  use the fact \cite{Nichuang} that  any entangling unitary combined with all 1-qubit gates yields all unitaries to conclude that we here obtain a universal set of gates through motion.

{\it Conclusions and Outlook.-} We have shown the fundamental result that through the controlled acceleration of qubits, such as an atom, universal 1-qubit operations (i.e., arbitrary rotations in the Bloch sphere) can be performed.  Controlled interactions of two suitably accelerated qubits, in principle, could yield all two-qubit gates and therefore universal quantum computing.  Although the high accelerations required are experimentally attainable in principle (see  \cite{ruso,BerryPh})  we discussed that an experiment is already within reach of quantum simulators in analogue systems.
It should be very interesting to determine optimal acceleration protocols, which may, e.g., involve oscillatory paths that employ resonance phenomena.

In addition, our finding here, namely that accelerations can induce arbitrary rotations in the Bloch sphere of an atomic qubit,  could also be useful in the reverse direction: by experimentally checking for subtle rotations in the Bloch sphere, of a detector in a suitable background, one may be able to better detect quantum effects that are due to the detector's acceleration, such as the Unruh effect, or even quantum effects that are due to curvature, due to the equivalence principle (see also \cite{Kerr,Collapse2}).

{\it Acknowledgements.-} The authors gratefully acknowledge support through NSERCÕs Banting, Discovery, PGSM, and Canada Research Chairs Programs. E.M.-M. thanks the partial support of the MICINN/MINECO Project No. FIS2011-29287. A.K. is grateful to the University of Queensland for the very kind hospitality during his sabbatical stay.

%

\end{document}